\documentstyle[psfig]{crckapb}

\newcommand{\tens}[1]{\sf #1}

\begin{opening}
\title{Analyzing X-ray variability by State Space Models}
\subtitle{Application to an EXOSAT AGN sample}                                         
\author{Michael K\"onig}
\author{R\"udiger Staubert}
\institute{Institut f\"ur Astronomie und Astrophysik - Astronomie,\\
               Universit\"at T\"ubingen, Germany}
\author{Jens Timmer}
\institute{Fakult\"at f\"ur Physik,
               Albert-Ludwigs-Universit\"at Freiburg, Germany}
\end{opening}

\begin{document}

\begin{abstract}
In recent years, autoregressive models have had a profound impact on the
description of astronomical time series as the observation of a stochastic
process. These methods have advantages compared with common Fourier
techniques concerning their inherent stationarity and physical
background. If autoregressive models are used, however, it has to be taken
into account that real data always contain observational noise often
obscuring the intrinsic time series of the object. We apply the technique
of a Linear State Space Model which explicitly models the noise of
astronomical data and allows to estimate the hidden autoregressive
process. As an example, we have analysed a sample of Active Galactic Nuclei
(AGN) observed with EXOSAT and found evidence for a relationship between
the relaxation timescale and the spectral hardness.
\end{abstract}

\section{Introduction and Mathematical Background}

Irregular X-ray variability is a common phenomenon of Active Galactic
Nuclei (AGN). The power spectra of the observed lightcurves exhibit an
`observational noise floor' at high frequencies, an increase of power
towards low frequencies ($1/f^\alpha$ behavior) and, most often in long term
lightcurves, a flat top at the low frequency end of the spectrum (McHardy
1988). As a commonly applied method the periodogram is used to estimate
part of the power spectrum with a slope $\alpha$ usually between 0 and 2
and a mean of about 1.5 (Lawrence and Papadakis 1993).

When the periodogram is computed, the window function (the Fourier
transform of the observational sampling) is convolved with to the true
spectrum of the source. This can produce artefacts in the power spectrum,
which make a proper interpretation more difficult (Papadakis and Lawrence
1995; Priestley 1992). In the case of AGN spectra the sidebands of the
window function yield spectral leakage to higher frequencies which will
cause the spectra to appear less steep and the spectral slope will be
underestimated (Deeter and Boynton 1982; Deeter 1984). Consequently a model
is required which operates in the time domain and avoids any misleading
systematic effects. As an alternative approach, we have used the Linear
State Space Model (LSSM). The LSSM is a generalization of the
autoregressive (AR) model invented by Yule (1927) to model the variability
of Wolf's sunspot numbers.

The AR model expresses the temporal correlations of the time series in
terms of a linear function of its past values plus a noise term and is
closely related to the dynamics of the
system. LSSMs generalize the AR processes by explicitly modelling
observational noise. The variable $x(t)$ that has to be estimated cannot be
observed directly since it is covered by observational noise $\eta(t)$. The
measured observational variables $y(t)$ do not generally agree with the
system variables $x(t)$. Thus an LSSM is defined with two equations, the
system or dynamical equation~(\ref{eq:sys}) and the observation
equation~(\ref{eq:obs}).
\begin{eqnarray}
 \vec{x}(t) & = & \tens{A} \, \vec{x}(t-1) +  \vec{\epsilon}(t) 
 \quad \vec{\epsilon}(t) \sim {\cal{N}} (0,\tens{Q}) \label{eq:sys} \\ 
 y(t) & = & \tens{C} \, \vec{x}(t) + \eta(t) 
 \quad \eta(t) \sim {\cal{N}}(0,R) \label{eq:obs}
\end{eqnarray}
This definition is a multivariate description, which means that the AR[p]
process is given as a $p$-dimensional AR process of order one, with a
matrix $\tens{A}$ that determines the dynamics. The matrix $\tens{C}$ maps
the unobservable dynamics to the 
observation. The terms $\vec{\epsilon}(t)$ and $\eta(t)$ represent the
Gaussian dynamical noise with covariance matrix $\tens{Q}$ and the Gaussian
observational noise with variance $R$, respectively. We have used the
Expectation-Maximization algorithm to estimate the dynamics and a
KS test to quantify the order of the hidden AR process
(Honerkamp 1993, K\"onig and Timmer 1997).

\section{The EXOSAT AGN sample} \label{exosat}

As the X-ray lightcurves from the X-ray satellite EXOSAT are the longest
existing observations of AGN, we have used individual observations longer
than 25\,ks to create an EXOSAT AGN timing sample (see table~1). We use the
lightcurve of the Seyfert galaxy NGC~5506 as an example for a brief
description of how to apply LSSM and interpret LSSM results. We have found
that the X-ray lightcurve of NGC~5506 can be well modelled with a LSSM
AR[1] model, since the residuals between the estimated AR[1] process and
the measured data are consistent with Gaussian white noise. The
corresponding dynamical parameter of the LSSM AR[1] fit corresponds to a
relaxation time of about 1.6 hours. Higher order LSSM AR[p] models cannot
improve the fits significantly. Therefore more complex models (with
additional relaxators and damped oscillators) are not needed to describe
the dynamics of the AGN lightcurve. In addition, an AR[1] model provides
all features occuring in AGN power spectra, especially the flat top at low
frequencies which cannot be described by the $1/f$ models (K\"onig and
Timmer 1997).

\begin{figure}
\centerline{\psfig{file=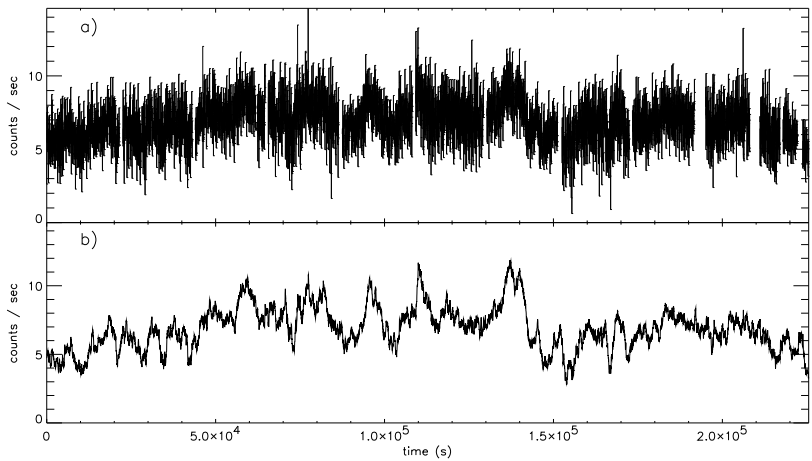,width=0.9\textwidth,clip=}}
\caption[ ]{a) EXOSAT ME X-ray lightcurve of NGC~5506 (30sec/bin,
Jan.~1986), b) Hidden AR[1]-process, estimated with the LSSM fit. Both
lightcurves are shown without error bars for clarity.}
\end{figure}

\section {Discussion} 

The physical interpretation of the discovered AR[1] dynamics is a
sto\-chastic superposition of single shots all decaying with the same
relaxation time.  The assumption of an exponentially decaying shot seems to
be reasonable as time-dependent Inverse Comptonisation (IC) models lead to
such a pulse profile (Liang and Nolan 1983).  In this scenario
inhomogeneities in the accretion flow produce single UV flares. These UV
photons are then upscattered in the Comptonising medium producing the
observed X-ray energy spectrum. Since harder X-ray photons need more
interactions in the Compton cloud and thus have experienced a longer stay
in the medium, they are temporally delayed with respect to the soft X-ray
photons. The LSSM fits of the EXOSAT AGN sample yield a relationship
between the spectral hardness (expressed by the slope $\Gamma$ of the
power-law fit of the photon spectrum), and the estimated relaxation time of
the AR process (table~1), which might be a fingerprint of the IC
process. The relation is derived to $\tau = 74890 - 38040 \cdot \Gamma$~sec
with a corresponding Kendalls tau value of 0.744 and a two-sided
significance level of $5.7 \cdot 10^{-5}$ for the hypothesis of
uncorrelated data.  Since the $\tau$--$\Gamma$ relation is independent of
the AGN type, a common physical mechanism for the spectral and temporal
behavior is needed for all kinds of AGN --- possibly in support of the
unifying model of Active Galactic Nuclei (Netzer 1990).

\begin{table}
\caption{LSSM fit results of the EXOSAT AGN sample}
{\tiny
\begin{tabular}[c]{cccccccc} \hline
AGN & \hspace*{-3mm}observation & $T_{\rm tot}^a$  & $L_{2-10keV}^b$      & $\tau^c$ &
    $KS_{\rm Test}^d$ & $\Gamma^e_{2-10keV}$ & ${\chi^{2}_{\rm{red}}}^f$  \\ 
    &             & ksec             & $\log_{10}$ erg/s & ksec     & \%
        &            &  \\ \hline 
NGC 4051 &337/85 &143.84 &41.182 &${2.38}^{+1.94}_{-0.74}$ &81.2 &${1.90}^{+0.04}_{-0.04}$ &0.99   \\
MKN 421 &338/84 &26.59 &44.439 &${3.41}^{+2.39}_{-1.00}$ &98.0 &${1.89}^{+0.13}_{-0.09}$ &1.93  \\
MCG 6-30-15 &28/86 &183.69 &42.659 &${3.63}^{+2.82}_{-1.11}$ &76.0 &${1.82}^{+0.14}_{-0.10}$ &-  \\
NGC 4593 &176/85 &32.49 &42.217 &${4.24}^{+2.67}_{-1.18}$ &93.5 &${1.82}^{+0.05}_{-0.04}$ &0.74 \\
NGC 5506 &24/86 &225.56 &42.777 &${5.89}^{+4.04}_{-1.71}$ &99.8 &${1.84}^{+0.04}_{-0.04}$ &0.95   \\
Fairall 9 &286/83 &30.60 &43.791 &${6.28}^{+4.81}_{-1.90}$ &83.6 &${1.87}^{+0.10}_{-0.10}$ &1.42   \\
NGC 4593 &9/86 &95.65 &42.613 &${8.72}^{+6.69}_{-2.64}$ &80.1 &${1.78}^{+0.05}_{-0.05}$ &1.01   \\
NGC 5548 &62/86 &85.88 &43.430 &${10.69}^{+7.45}_{-3.11}$ &85.8 &${1.61}^{+0.03}_{-0.03}$ &1.47  \\
3C 120 &228/83 &44.61 &43.896 &${11.34}^{+7.42}_{-3.22}$ &84.9 &${1.70}^{+0.04}_{-0.04}$ &0.74  \\
Cen A &44/84 &44.22 &42.147 &${13.29}^{+11.48}_{-4.21}$ &94.4 &${1.75}^{+0.08}_{-0.08}$ &1.28  \\
3C 120 &276/84 &44.52 &44.029 &${13.30}^{+11.50}_{-4.22}$ &97.0 &${1.71}^{+0.03}_{-0.02}$ &1.39   \\
NGC 5548 &19/86 &63.75 &43.545 &${13.54}^{+12.11}_{-4.35}$ &95.9 &${1.64}^{+0.03}_{-0.03}$ &0.83 \\
NGC 1068 &8/85 &54.35 &41.262 &${16.05}^{+8.43}_{-4.11}$ &83.1 &${1.68}^{+0.07}_{-0.05}$ &0.71   \\
3C 273 &17/86 &145.20 &46.402 &${18.02}^{+11.35}_{-5.03}$ &98.4 &${1.52}^{+0.02}_{-0.02}$ &0.90   \\
NGC 4151 &192/83 &86.66 &42.327 &${28.54}^{+32.43}_{-9.91}$ &75.0 &${1.23}^{+0.11}_{-0.11}$ &1.31   \\
NGC 4151 &27/85 &94.45 &42.393 &${31.77}^{+46.14}_{-11.82}$ &90.7 &${1.53}^{+0.11}_{-0.10}$ &0.60  \\
\hline
\end{tabular} 
$^a$total observation time, $^b$X-ray luminosity in the 2--10 keV energy
range , $^c$AR relaxation time, $^d$Kolmogorov-Smirnov test on white noise
residuals, $^e$slope of the power law fit of the photon spectrum in the
2--10 keV energy range, $^f$reduced $\chi^{2}$ of the photon spectrum fit

}
\end{table}


\begin{thebibliography} {999}
\bibitem []{}
Deeter J.E., 1984, ApJ 281, 482-491
\bibitem []{}
Deeter J.E., Boyton P.E., 1982, ApJ 261, 337-350
\bibitem []{}
Honerkamp J., 1993, Stochastic Dynamical Systems, VCH Publ. New York, Weinheim 
\bibitem []{}
K\"onig M., Timmer J., 1997, A\&A, accepted for publication
\bibitem []{}
Lawrence A., Papadakis P., 1993, ApJ Suppl. 414, 85
\bibitem []{}
Liang E.P., Nolan P.L., 1983, Space Sci. Review 38, 353
\bibitem []{}
McHardy I., Czerny B., 1987, Nature 325, 696
\bibitem []{}
Netzer H., 1990, Saas-Fee Advanced Course 20 on Active Galactic Nuclei, eds. R.D. Blandford, H.Netzer, L.Woltjer, 57-160
\bibitem []{}
Papadakis I.E., Lawrence A., 1995, MNRAS 272, 161
\bibitem []{}
Priestley M.B., 1992, Spectral Analysis and Time Series, San Diego, Academic Press
\bibitem []{}
Scargle J.D., 1981, ApJ Suppl. 45, 1
\bibitem []{}
Yule G., 1927, Phil. Trans. R. Soc. A 226, 267 
\end {thebibliography}
\end{document}